# Accurate transfer of individual nanoparticles onto single photonic nanostructures

J. Redolat[1], M. Camarena-Pérez[1], A. Griol[1], M. Kovylina[1], A. Xomalis[2,3], J. J. Baumberg[2], A. Martínez [1*] and E. Pinilla-Cienfuegos [1*]

[1]Nanophotonics Technology Center, Universitat Politècnica de València, Valencia E46022, Spain
[2]NanoPhotonics Centre, Cavendish Laboratory, Department of Physics, JJ Thompson Avenue, University of Cambridge, Cambridge CB3 0HE, United Kingdom
[3]Empa, Swiss Federal Laboratories for Materials Science and Technology, Laboratory for Mechanics of Materials and Nanostructures, Thun, Switzerland

*Corresponding authors: Elena Pinilla Cienfuegos & Alejandro Martínez

**Abstract:** Controlled integration of metallic nanoparticles (NPs) onto photonic nanostructures enables realization of complex devices for extreme light confinement and enhanced light-matter interaction. This can be achieved combining Nanoparticle-on-Mirror (NPoM) nanocavities with the light manipulation capabilities of micron-scale metallic antennas and/or photonic integrated waveguides. However, metallic nanoparticles are usually deposited via drop-casting, which prevents their accurate positioning. Here we present a methodology for precise transfer and positioning of individual NPs onto different photonic nanostructures. The method is based on soft lithography printing that employs elastomeric stamp-assisted transfer of individual NPs onto a single nanostructure. It can also parallel imprint many individual NPs with high throughput and accuracy in a single step. Raman spectroscopy confirms enhanced light-matter interactions in the resulting NPoM-based devices. Our method mixes *top-down* and *bottom-up* nanofabrication techniques and shows the potential of building complex photonic nanodevices for applications ranging from enhanced sensing and spectroscopy to signal processing.

**Introduction**

Nanoparticle-on-a-mirror (NPoM) cavities offer extreme light confinement within nm-scale gaps.[1–4] Essentially a NPoM (often termed patch antenna or metal-insulator-metal MIM waveguide) is formed by depositing a metallic NP on top of a metallic mirror covered by a self-assembled molecular monolayer (SAM) so that under suitable illumination, light is confined in the nm-scale gap separating the metal surfaces. For a wide range of applications, such as absorbing elements,[5] NPoM are assembled over large areas so NPs are deposited on metal mirrors without requiring any position accuracy. Conventionally this can be done via drop-casting of NPs onto metallic surfaces covered by a SAM, which has become the most straightforward way to produce nm-scale cavities with high accuracy. However, some applications – such as molecular frequency upconversion [6,7] may require the integration of a single NP with a µm-scale antenna instead of a metal mirror.[8,9] Here, depositing single NPs onto nanostructures with high positional accuracy is of major importance. This is because for instance focussing mid-infrared light on an antenna localises the optical field in specific locations, and the NPoMs have to be placed in these locations to maximise nonlinear interactions with visible/near IR light.

Current NPs printing techniques developed for the controlled imprint of NPs on surfaces are based on colloidal self-assembly methods or utilising capillary forces within polymeric templates.[10–14] Single NP positioning onto individual lithographed nanostructures is challenging. So far, single NP positioning can be obtained via laser printing onto glass [15,16] (not metals) or by Atomic Force Microscopy (AFM)[17,18] but these are complex, expensive and slow (serial) methods.

In this work, we introduce a large-scale method for the accurate delivery of single NPs on complex photonic nanostructures (such as µm-scale metallic antennas antennas or integrated waveguides) based on a stamp-assisted soft lithography method. The main advantage of soft lithography is that it is a parallel nanoprinting technique that provides high-throughput and high simplicity together with the possibility of the precise transfer of multiple individual NPs onto different planar and non-planar nanostructures. A key point is that it can be used for both single-step positioning of multiple individual NPs onto an array of antennas as well as for the positioning of a single NP onto an individual photonic nanostructure. Our approach is cost-effective and a robust nanolithography methodology that does not require complex (and expensive) instrumentation. We validate our method by performing transfer of NPs onto different kinds of photonic structures, such as µm-scale plasmonic antennas and integrated $Si_3N_4$ waveguides. In addition, we show enhanced light-matter interaction in representative resulting devices using surface-enhanced Raman spectroscopy (SERS) measurements.

Our method (Figure 1) is based on lithographically-controlled wetting (LCW) that provides a route for in-situ fabrication of NPoM cavities based on elastomeric stamps, typically made of poly(dimethylsiloxane) (PDMS) (Figure 1i).[19] The PDMS stamps are designed according to the target photonic structures with the aim to positioning the NPs at each single antenna or waveguide (Figure 1ii). The LCW consists of the stamp-assisted deposition of a soluble material from a solution (or in our case, a colloidal suspension of Au NPs in water). To facilitate the delivery of the NPs, the lithographed motifs of the sample are functionalized with a self-assembled monolayer (SAM) that improves the affinity between the NP and the substrate (Figure 1iii). As the stamp is placed in contact with a colloidal suspension on a surface, the capillary forces drive the liquid to distribute only under the protrusions of the stamp producing an array of menisci (Figure 1iv). The NPs are transferred in a pattern defined by the topography of the stamp, with a minimum feature size as small as the meniscus formed at the stamp protrusion (i.e., pillar, Figure 1v). The NP is locally trapped in the meniscus, and due to the chemical affinity with the functionalized surface, the NP transfers to the surface. The stamp is then lifted-off and, as the solvent evaporates, individual NPs are patterned on the surface with the same length scale as that of the stamp (Figure 1vi).

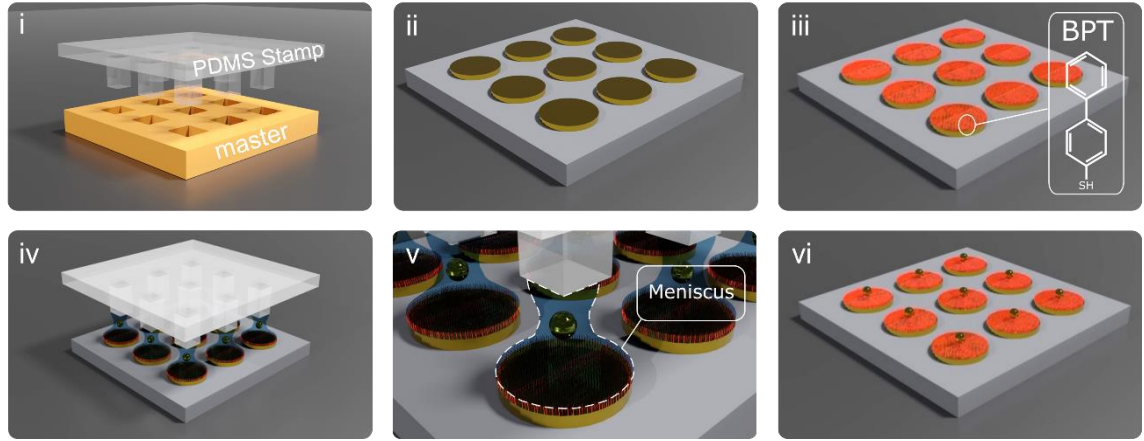

*Figure 1.* **Schematic parallel printing method**: i. PDMS stamp fabrication by cast moulding; ii. Fabrication of the photonic micro-nanostructure; iii. Sample functionalization with biphenyl-4-thiol (BPT) SAM; iv. NP printing process with localised menisci.; v. Meniscus formation between the stamp protrusion and the functionalized lithographed sample. vi) Each NP is trapped and guided by the meniscus to the antenna array; vi. Lift off PDMS stamp leaving NPs attached to the sample.

The concentration of the NP colloidal suspension, the affinity between stamp-solution-substrate, and the applied pressure between stamp and the target nanostructure are parameters that can be modified to deliver different patterns of the same components without modifying the stamp features. Moreover, the flexibility of the PDMS stamp and the ability to achieve conformal, nanometric level contact between the stamp and the substrate are both advantageous for printing over nanometric and micrometric photonic antennas such as disks, bow-ties, waveguides, or even curved substrates. Moreover, the PDMS stamp shows higher hydrophobicity (contact angle ~ 100˚, see SI 1) than the functionalized substrates (contact angle of ~ 80˚ for BPT-Au antennas, see SI 2) which facilitates the meniscus formation and, therefore, the suitable conditions for trapping/transferring NPs.

One of the crucial parameters required for the deposition of single NPs onto specific locations onto the antennas array is the precise alignment between the stamp and the array substrate. To provide this, a µ-printing device comprising pressure monitoring is constructed to deliver sub-micrometric accuracy and high reproducibility with precise control at every step of the stamping method.

Here, an optical microscope (O1) is mounted on a XYZ translational stage allowing coarse alignment and focusing on the z-axis (Figure 2). O1 allows µm-scale precision with the aid of a digital camera and white illumination launched through via an optical fiber. A 4D stage (XYZѲ) aligned with O1, is used for the nm-scale sample positioning. A force sensitive resistor (FSR) is placed on top of the translational stage for pressure monitoring while the sample lay on top. A second translational 3D stage is used to place the stamp. The PDMS stamp is mounted on a glass slide, attached with transparent sticky tape to allow monitoring from the top (O1). The glass slide is fixed to the translational stage with a 3D printed piece purpose-designed for this configuration. This custom-build holder can be screwed on the 3D stage to adapt the position of the PDMS stamp over different samples. In addition, two optical microscopes (O2 and O3) mounted on adjustable height posts allow monitoring along the z-axis. They are connected to a PC via USB and are situated at 90° to facilitate the correct alignment of the system stamp/substrate. They allow to real time monitoring to enable feedback and quantification.

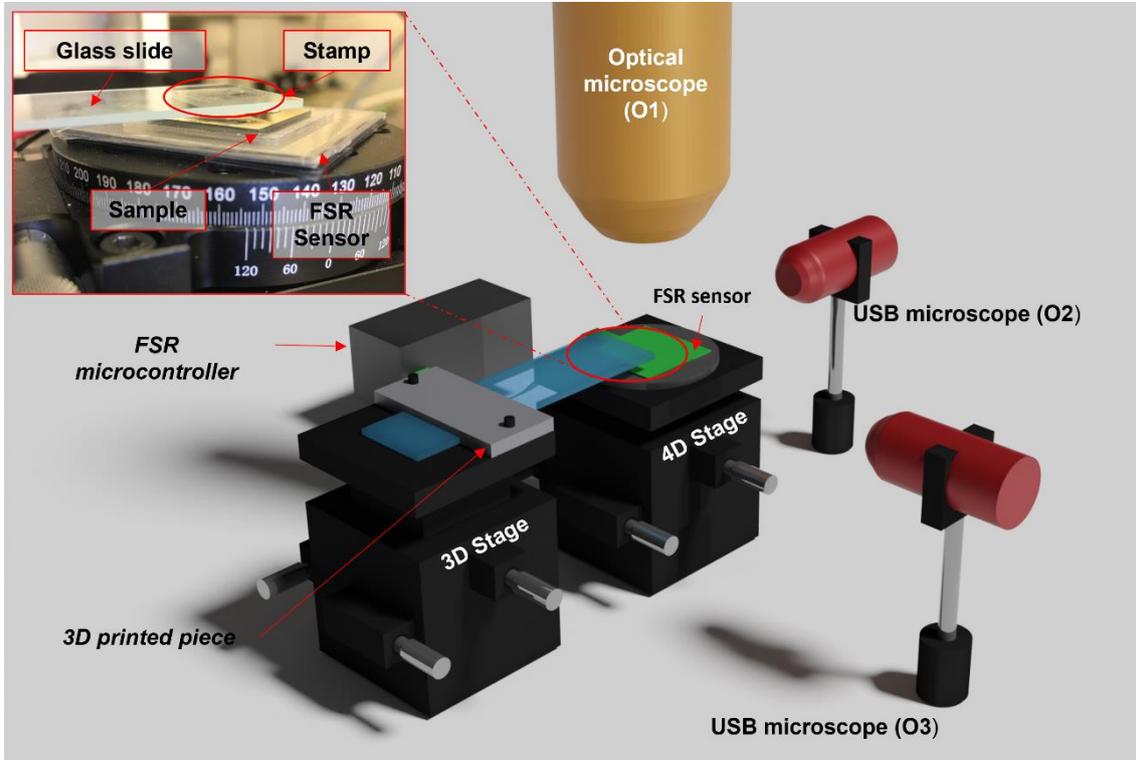

*Figure 2.* 3D schematic of the nanoimprint setup. Inset shows a photo of the stamp, sample positioning and force sensitive resistor (FSR).

The printing experiment is performed as follows: First, 100 µL of NP solution drop casted onto the stamp for about 2 minutes. Then, excess solution is removed with a tissue from the side of the stamp (not from top). The glass slide with the 'wet' stamp is then fixed to the 3D stage and rapidly aligned on top of the nanostructured sample with the micropositioning mountings. Manual alignment is optically controlled via O1. Once the stamp is correctly aligned with the sample, it is approached until it is pressed with a force of ~2 N. Finally, the stamp if lifted-off the sample. For the imprinting to succeed, this whole process must take less than ~2 minutes - otherwise, the solvent evaporates, and the transfer yield dramatically decreases.

To prove that this yields a single-step large-scale nanopositioning method for the parallel transfer of individual gold NPs on multiple photonic structures in one single step, we use three samples (Disk 1, Disk 2 and Disk 3). The samples consist of 4 x 4 arrays of gold resonators disks on a silicon substrate. Each array was formed by 24 x 24 Au disk antennas (sample Disk1 = 2304 disks) and 25 x 25 disks (samples Disk 2 and Disk 3 = 2400 disks) of diameter $\Phi_D$ = 6 µm, with 1 µm separation (P = 8 µm pitch) and 120 nm height (Figure 3a). A Si master was fabricated of identical sample dimensions for producing the polymeric stamps (see Methods section). Hollow squares were lithographically fabricated with two different widths (WS) of W1$_S$ = 1 μm in the case of Stamp1 and W2$_S$ =2 μm for Stamp2, and depths of 550 nm, using 8 μm pitch in both cases to match the target samples (Disk1 and Disk2,Figure 3b). The polymeric stamps were then prepared by cast moulding over the fabricated master substrates (Methods section, (Figure 3c). Two PDMS stamps were patterned of square relief features with the same number of columns as the patterned gold disks samples and same pitch to ensure a perfect match. In the first case (Stamp1), square 24 x 24 columns of side W1$_S$ = 1 μm and height of H$_S$ = 550 nm (P = 8 μm pitch), while the second case (Stamp2) uses 25 x 25 columns of side W2$_S$ = 2 μm and height of H$_S$ = 550 nm (P = 8 μm pitch) (Figure 3d). The gold disk antennas array was functionalized with a biphenyl-4-thiol (BPT) SAM for effective NP transfer in both cases. Once the sample was functionalised, spherical Au-NPs of 60 nm diameter were imprinted in single-step.

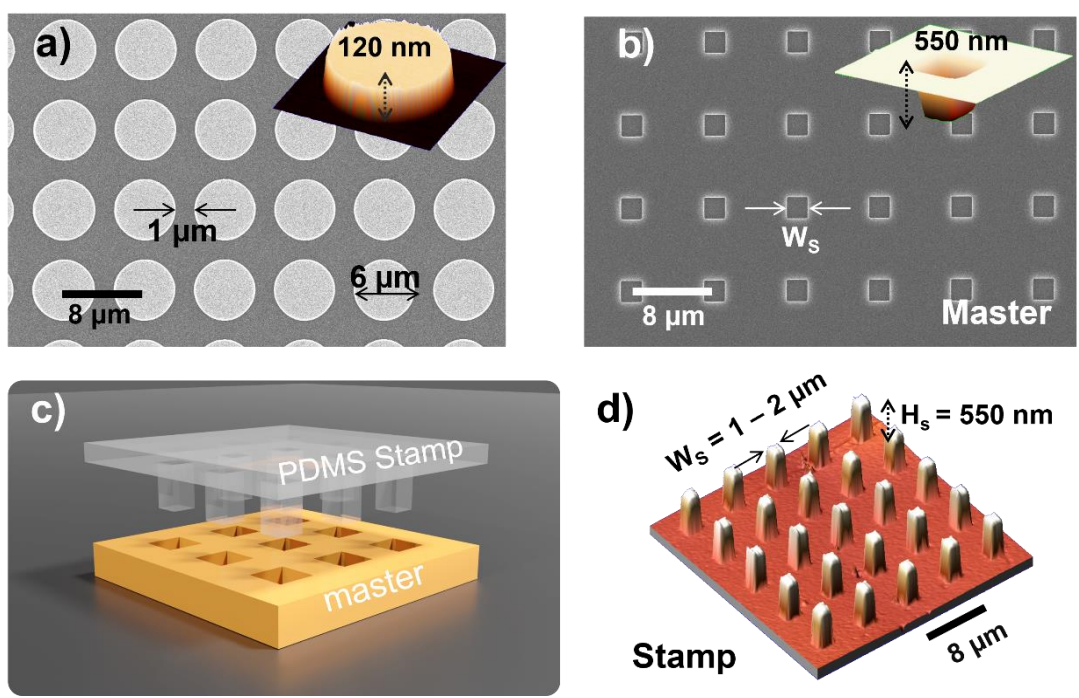

*Figure 3. **Nanoparticle transfer substrates and stamp master.** a)* SEM image of Au resonator disk array of 6 μm diameter, 1 μm separation and 8 μm pitch. Au disks are patterned onto a Si substrate. Inset shows AFM measurement of single Au disk of 120 nm height. b) SEM image of master on Si used for the PDMS stamp moulding. Hollow squares have width W1$_S$ = 1 μm (Stamp1) and W2$_S$ =2 μm (Stamp2), and depths of 550 nm, 8 μm pitch for both. Inset shows AFM measurement of a single pit giving 550 nm depth. c) Sketch of PDMS stamp fabrication from the Si master. d) 3D AFM image of a fabricated PDMS stamp indicating its geometrical dimensions.

Imprint experiments were performed with the two designed stamps and two functionalized samples of Au disk arrays (BPT-Au disks, Disk1 and Disk2). In addition, a regular drop-casting experiment was performed depositing the same concentration of colloidal dispersion directly onto a third BPT-Au disk sample (Disk3) to compare the yield of the transfer method. We present statistics of the first experiment (Stamp1) in **Table 1**, and of the second experiment (Stamp2) in **Table 2**, accordingly. For each array the number of disks with single NPs (1 NP), two (2 NP), three NPs (3 NP) and clusters (>3 NP) as well as the percentage yield (considering the total number of disks: 576 disks for each array (A1 - A4) of Disk1 sample, and 625 for each array (A1 - A4) of Disk2 sample).

| Sample | Arrays (24 x 24) | | 1 NP | 2 NP | 3 NP | >3 NP |
|---|---|---|---|---|---|---|
| Disk 1 | A1 | Disks | 87 | 60 | 31 | 147 |
| | | Yield (%) | 15.1 | 10.4 | 5.4 | 25.5 |
| | A2 | Disks | 90 | 98 | 88 | 203 |
| | | Yield (%) | 15.8 | 17.0 | 15.3 | 35.1 |
| | A3 | Disks | 129 | 41 | 21 | 38 |
| | | Yield (%) | 22.4 | 7.1 | 3.6 | 6.6 |
| | A4 | Disks | 164 | 115 | 53 | 43 |
| | | Yield (%) | 28.5 | 17.0 | 9.2 | 7.5 |

*Table 1.* Nanoparticle positioning statistics for Stamp1 ($W1_S$ = 1 µm) transfer onto Sample Disk1. The Highest yield of 28% is highlighted in green.

| Sample | Array (25 x 25) | | 1 NP | 2 NP | 3 NP | >3 NP |
|---|---|---|---|---|---|---|
| Disk 2 | A1 | Disks | 19 | 4 | 2 | 2 |
| | | Yield (%) | 3.0 | 0.6 | 0.3 | 0.3 |
| | A2 | Disks | 116 | 78 | 52 | 135 |
| | | Yield (%) | 18.6 | 12.5 | 8.3 | 21.6 |
| | A3 | Disks | 67 | 57 | 37 | 376 |
| | | Yield (%) | 10.7 | 9.1 | 5.9 | 30.2 |
| | A4 | Disks | 20 | 5 | 1 | 3 |
| | | Yield (%) | 3.2 | 0.8 | 0.2 | 0.5 |

*Table 2.* Nanoparticle positioning statistics for Stamp2 ($W2_S$ = 2 µm) transfer onto Sample Disk2.

In sample Disk1 (with Stamp1, $W1_S$ = 1 µm), we find stamping yields of 15-28 % for single NP transfer while for sample Disk 2 this varies between 3-19 %. The use of a stamp with narrower pillars (550 nm height x 1 µm side columns) leads to a better single NP transfer. The nanopositioning of 60 nm Au-NPs is thus more efficient with the $W1_S$ = 1 µm PDMS pillars. We note that both cases result in much higher NP transfer yield than simple drop-casting (~1%), as expected. Finally, we emphasize that these experiments are performed in a single-step and the yield can be further improved by repeating the transfer step. Moreover, the elastomeric PDMS stamps are mechanically and chemically stable allowing reuse >50 times over several months without noticeable degradation in performance.

To better quantify our single-step printing, we show large-scale printing of single NPs onto disk antenna arrays (Figure 4). Examples of single NP positioning onto the disk antennas is shown for the Disk1 sample (Figure 4a). An area of 7 x 5 Au disks from Disk 2 sample is shown before and after the NP transfer (Figure 4b, c). Every disk contains a single NP placed in the same position over the disk (marked with red arrows). Only in few cases, we find two NPs transfer in one disk. To confirm the positioning and size of the transferred NPs we show AFM images of Disk2 before and after the transfer (Figure 4d-f). It is possible using the µ-printing device to laterally control the stamp onto nanometric structures within 650 nm (sub-micrometric resolution), so able to deliver single NPs onto individual structures. This evidence the capability of controlled single NP printing onto different complex nanostructures.

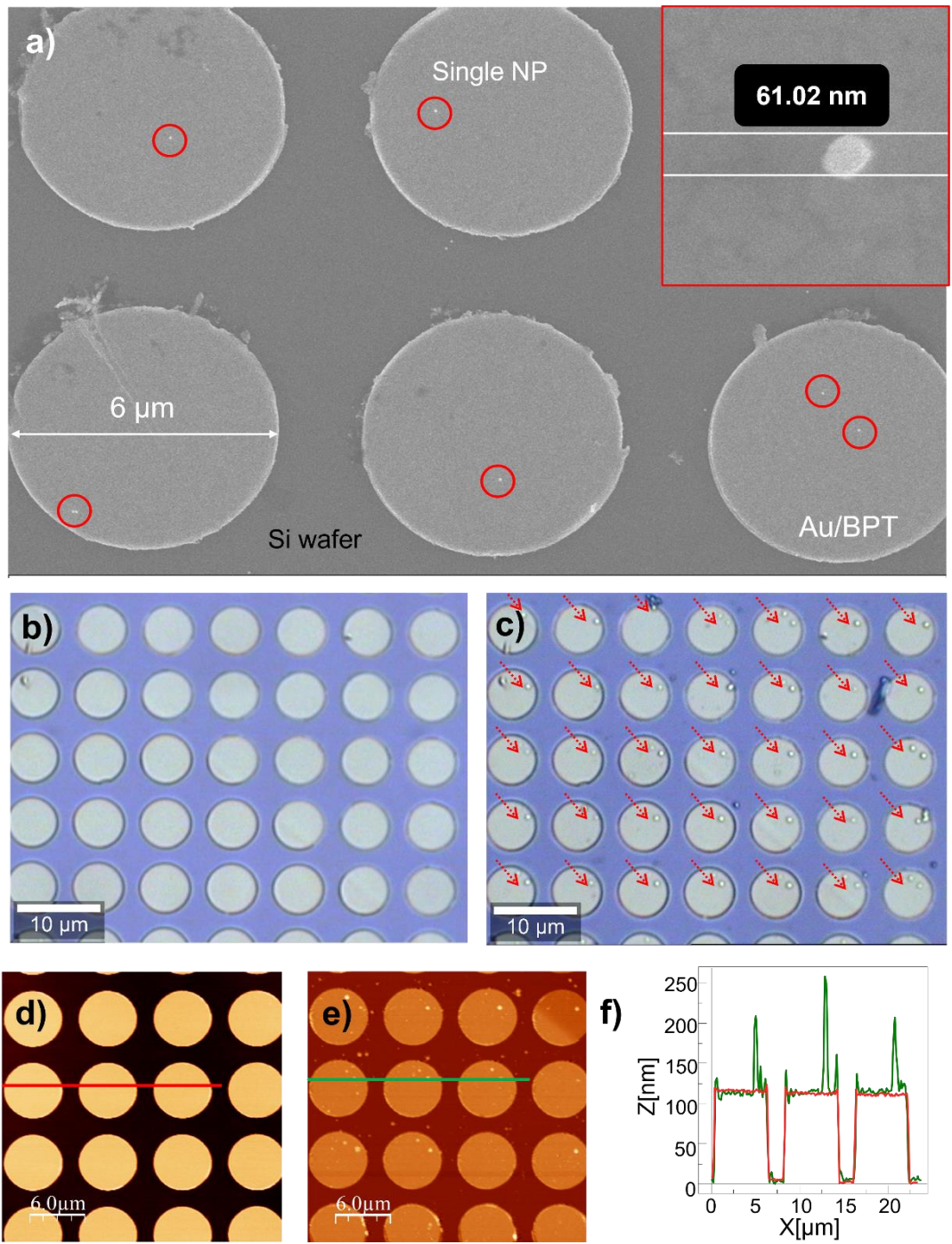

*Figure 4. **High fidelity single-step large-scale single nanoparticle (NP) transfer method**. a) SEM image of 60 nm NPs transferred individually with $W1S = 1$ µm stamp. b) Optical microscope image of Au disk array of 6 µm diameter fabricated on a Si wafer before transfer of NPs with $W2S = 2$ µm stamp. c) Optical microscope image of the same area of functionalized Au/BPT disks after single-step NP transfer with $W2S = 2$ µm stamp. AFM measurements of the same area before (d) and after (e) the NP transfer. f) AFM profile for three Au/BPT disks with single NPs printed on top.*

To evidence this transfer to standard photonic devices, we show the transfer method onto a metallic µm-scale disk antenna fabricated interfacing a $Si_3N_4$ waveguide. This photonic structure has been proposed as a promising platform for on-chip SERS sensing [7,9] as it delivers high in- and out-coupling signal efficiencies. The aim here is to place a single NP on the disk antenna to form a so-called Nanoparticle-on-Resonator (NPoR)[8] desing, with the waveguide used to couple light efficiently in and out of the nanocavity. Using the same transfer strategy, we deliver single NPs onto Au disks at the waveguide end, for several disk diameters: i) 6 µm , ii) 5 µm, and iii) 4 µm (Figure 5a). In this case, stamps with 1 µm x 1 µm pillars were used to form smaller menisci and give better yields. Single 60 nm Au-NPs were successfully positioned on all the Au/BPT functionalized disks (Figure 5b). To validate the presence of the BPT-SAM in the fabricated NPoR photonic structure, we perform SERS experiments with free-space excitation and collection from above (Figure 5c). The SERS spectrum of BPT (case ii of 5 µm disk) is obtained (Figure 5d) for pump wavelengths of 633 nm (red) and 785 nm (blue) which shows enhanced vibrational BPT signatures due to the elevated near-field of the NPoR geometry.

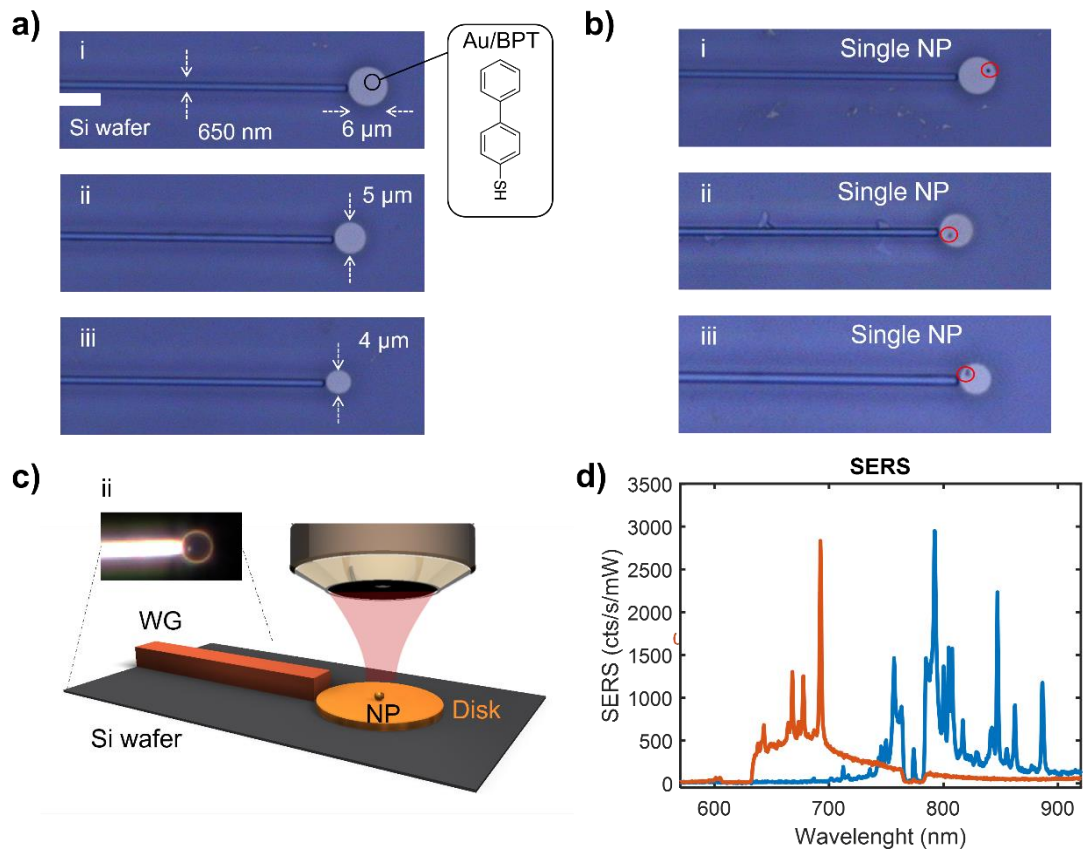

*Figure 5. **Single nanoparticle (NP) stamping onto Nanoparticle-on-Resonators (NPoR) coupled with standard $Si_3N_4$ waveguides.** A) Optical microscope images of BPT-functionalized Au disks of different diameters: i: 6 µm, ii: 5 µm and iii: 4 µm at the end of 650 nm wide $Si_3N_4$ waveguides (WG). B) Optical microscope images of same structures with NPs of 60 nm diameter printed on top. c) Scheme of NPoR interfacing a WG and free-space SERS characterization. Inset shows dark-field image of case ii. d) BPT SERS spectrum for pump wavelengths of 633 nm (red) and 785 nm (blue).*

Further, single Au NP positioning was also achieved onto $Si_3N_4$ waveguides functionalized with (3-aminopropyl)triethoxysilane (APTES) (Figure 6a-b). In these cases, the transfer found to be more challenging, not only for the difficulty in the stamp alignment but also because the meniscus formation is less stable in this type of nanostructures (better in round-shaped antennas). However, after two or three transfer repeats, single NP transfer is achieved in all cases.

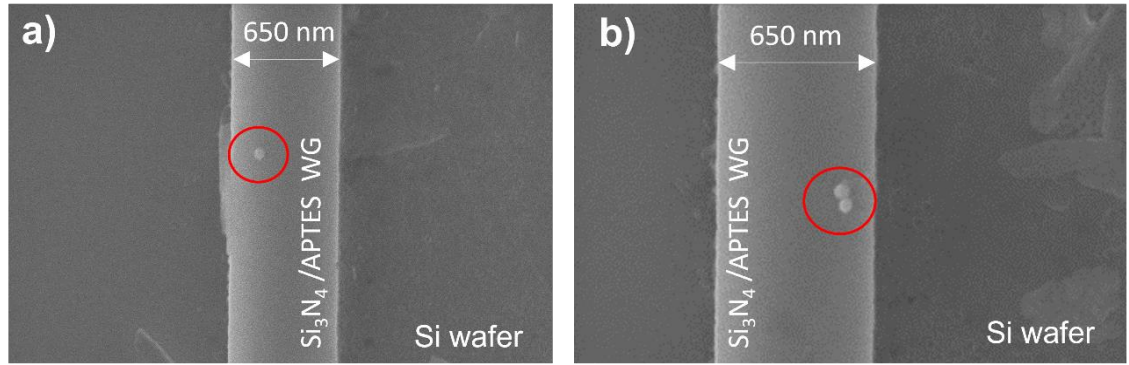

*Figure 6. **Single nanoparticle (NP) transfer onto Si$_3$N$_4$ waveguides (WGs).** SEM image of a) single and b) double Au NPs transferred onto a Si$_3$N$_4$ WG pre-functionalized with 3-aminopropyl)triethoxysilane APTES.*

**Conclusion**

We developed a reproducible, single-step, and cost-effective method for the controlled nanopositioning of single NPs for both parallel printing and single positioning of individual NPs onto standard lithographically fabricated photonic nanostructures with sub-micron accuracy in a single step. Taking advantage of the capillary forces in elastomeric stamps and utilising a custom-built µ-positioning device, we achieve a single-step NP transfer yield of up to 28%. In addition, the methodology is utilised to transfer NPs to more complex photonic structure geometries such as metallic disk antennae and integrated waveguides improving not only the drop-casting yield but also gaining in the control of the NP positioning over the micro-nanostructures. This large-scale approach opens the path towards deterministic NPoM fabricated cavities on nanophotonic structures for advanced spectroscopic architectures on-a-chip.

**Materials and Methods**

*µ-printing device*

Manual micropositioners have a maximum travel of 13 mm at XY along each XY axis, 10 mm along Z axis with sub-micron resolution, the goniometer (Ө) is used for sample rotation. FSR: Force Sensitive Resistor. Squared sensing area of 5 x 5 mm (Thickness of 0.45 mm), actuation force 0.1N and sensitivity range to 10N. For the electronic design, a Nano Arduino based on the ATmega328P microcontroller is used.

*STAMP and MASTER fabrication*

PMDS stamps were prepared by replica molding of the fabricated master. The PDMS stamps fabrication was done employing the kit named: "Kit SiliconElastomer Sylgard 18". Sylgard 184 is a bicomponent system for the fabrication of silicone stamps that is formed by a base and a curling agent. Once the proper mixture was prepared, the PDMS stamps were cured at 90°C for 45 minutes and peeled off the master.

The silicon master, consisting of a periodic array of squared wells, was fabricated using a standard direct writing process based on electron beam lithography. The fabrication was carried out on standard silicon samples (resistivity ρ ~$1^{-10}$ W.cm$^{-1}$, with a lightly p-doping of ~$10^{15}$ cm$^{-3}$). The fabrication process is based on an electron beam direct writing process performed on a coated 100 nm Poly(methyl 2-methylpropenoate) (PMMA) resist film. The mentioned electron beam exposure, performed with a Raith150 tool, was optimized in order to reach the required dimensions employing an acceleration voltage of 10 KeV and an aperture size of 30 µm. After developing the PMMA resist, the resist patterns were transferred into the Silicon samples employing an optimized Inductively Coupled Plasma- Reactive Ion Etching (ICP-RIE) process with fluoride gases (SF6 and CF4). Finally, the master was Al-coated to facilitate the peel off the PDMS stamp from the master. Otherwise, the PDMS polymer can break in the peel off step due to adhesion to the micrometric lithographed motifs. Once fabricated, the master can be reused multiple times as well as the stamps.

*BPT functionalization*

Biphenyl-4-Thiol (BPT, 97%) molecule was purchased from *Merck-Sigma Aldrich*. First, piranha solution ($H_2SO_4/H_2O_2$, 1:1) was used for glass cleaning. BPT SAMs were prepared by dipping the substrates in 1 mM BPT in ethanol (absolute, reagent grade) for 14h. Finally, the sample was sonicated in ethanol for 3 minutes, rinsed with ethanol, and dried under $N_2$ stream.

The quality of the SAMs was evaluated by Advancing-Receding Contact Angle measurements in a Ramé-hart automatized goniometer and by AFM imaging.

*Drop casting Au-NP deposition*

The drop casting was performed delivering a 10 µl drop of 60 nm Au-NP solution onto the Disk3 sample, left for 5 minutes and then rinsed with Mili-Q water. Finally, the substrate was dried under $N_2$ stream. Citrate-capped 60 nm Au NPs were purchased from Nanopartz™.

*APTES Functionalization*

3 – aminopropyl triethoxysilane (APTES, 99%) molecules and solvents were purchased from *Merck-Sigma Aldrich* and used without previous purification. APTES SAMs were prepared by dipping the substrates in 1 mM APTES in ethanol (absolute, reagent grade) for 45 minutes, then rinsed with ethanol and finally dried under $N_2$ stream.

*AFM imaging*

Alpha300 RA (Raman-AFM) from *WITec* was employed for the AFM sample characterization. All measurements were performed in AC mode. Sharp silicon probes without coating (K ~42 N/m, $f_0$ ~320kHz) were purchased from PPP-NCH (Nanosensors). All AFM images were processed with WSxM software from Nanotec Electrónica S.L.[20]

**Acknowledgements**

We acknowledge support from European Research Council (ERC) under Horizon 2020 research and innovation programme THOR H2020-EU-8290 (Grant Agreement No. 829067) and PICOFORCE (Grant Agreement No. 883703). This work was also supported by funding from Generalitat Valenciana (Grants No. PROMETEO/2019/123, BEST/2020/178 and IDIFEDER/2021/061) and the Spanish Ministry of Science and Innovation (ICTS-2017-28-UPV-9 and PGC2018-094490-BC22). E.P.-C gratefully acknowledges funding from Generalitat Valenciana (Grant No. SEJIGENT/2021/039). A.X. acknowledges support from the Empa internal funding scheme (IRC 2021).